\documentclass[10pt,journal,compsoc]{IEEEtran}
\ifCLASSOPTIONcompsoc
  \usepackage[nocompress]{cite}
\else
  \usepackage{cite}
\fi
\ifCLASSINFOpdf
   \usepackage[pdftex]{graphicx}
   \DeclareGraphicsExtensions{.pdf,.jpeg,.png}
\else
\fi
\usepackage{subfigure}
\usepackage{amssymb}

\usepackage{xcolor}

\usepackage[linesnumbered,ruled]{algorithm2e}

\usepackage{url}

\begin{document}
%
\title{DOMINO: Visual Causal Reasoning with Time-Dependent Phenomena}
%
%
%
%

\author{Jun~Wang,~and~Klaus~Mueller,~\IEEEmembership{Senior~Member,~IEEE}
\IEEEcompsocitemizethanks{\IEEEcompsocthanksitem Jun Wang and Klaus Mueller are with the Visual Analytics and Imaging Lab at the Computer Science Department, Stony Brook University, Stony Brook, NY. E-mail: \{junwang2, mueller\}@cs.stonybrook.com.}
\thanks{Manuscript received April 19, 2005; revised August 26, 2015.}}

%
%

\markboth{IEEE TRANSACTIONS ON VISUALIZATION AND COMPUTER GRAPHICS (ACCEPTED 2022)}%
{Shell \MakeLowercase{\textit{et al.}}: Bare Demo of IEEEtran.cls for Computer Society Journals}
%



\IEEEtitleabstractindextext{%
\begin{abstract}
Current work on using visual analytics to determine causal relations among variables has mostly been based on the concept of counterfactuals. As such the derived static causal networks do not take into account the effect of time as an indicator. However, knowing the time delay of a causal relation can be crucial as it instructs how and when actions should be taken. Yet, similar to static causality, deriving causal relations from observational time-series data, as opposed to designed experiments, is not a straightforward process. It can greatly benefit from human insight to break ties and resolve errors.  We hence propose a set of visual analytics methods that allow humans to participate in the discovery of causal relations associated with windows of time delay. Specifically, we leverage a well-established method, logic-based causality, to enable analysts to test the significance of potential causes and measure their influences toward a certain effect. Furthermore, since an effect can be a cause of other effects, we allow users to  aggregate different temporal cause-effect relations found with our method into a visual flow diagram to enable the discovery of temporal causal networks. To demonstrate the effectiveness of our methods we constructed a prototype system named DOMINO and showcase it via a number of case studies using real-world datasets. Finally, we also used DOMINO to conduct several evaluations with human analysts from different science domains in order to gain feedback on the utility of our system in practical scenarios.
\end{abstract}

}

\maketitle

\IEEEdisplaynontitleabstractindextext

%
\IEEEpeerreviewmaketitle

\IEEEraisesectionheading{\section{Introduction}\label{sec:introduction}}

%
%
%
%
\IEEEPARstart{R}{evealing} the causal explanations of an observed phenomenon is one of the ultimate goals of data analysts, yet it is one of the most difficult tasks in science. The advantage of knowing causality, rather than just correlation, is that the former provides much clearer guidance in predicting the effects of actions. To tackle this challenge, modern statistical theories on causal inference have been well established following the illuminating work of Pearl \cite{Pearl2000} and Spirtes et al. \cite{Spirtes2000}.  
Generalized visual analytics frameworks leveraging these theories have also been presented recently \cite{Wang2016,Wang2017}, providing interactive utilities by which humans can apply their domain expertise to aid the causal inferencing.

While knowing that a causal relation exists is enlightening, knowing \textit{when} the change will occur can also be crucial, as it instructs how and when actions should be taken. For example, knowing the timing of biological processes will allow us to intervene properly to prevent disease; knowing the causes that drive the price of a stock in the stock market will enable profitable trading; knowing that secondhand smoking causes lung cancer in 10 years may motivate people to kick the habit and lead to legislation that prohibits public smoking, while on the other hand, people would be far less worried if the time delay was 90 years. This fine but powerful nuance of time is at the very root of causality.

Although theoretical tools analyzing the time factor in causality, e.g., \textit{Granger causality} \cite{Granger1969}, \textit{Dynamic Bayesian networks} \cite{Friedman1998}, and \textit{logic-based causality} \cite{Kleinberg2009,Kleinberg2011}, are widely adopted in scientific research, there are few, if any, interactive visual analytics tools that support domain users in these analytical tasks. Human analysts must resort to simple text-based editors to identify important phenomena and set up parameters for hypotheses evaluations. This might be feasible when testing a small number of relations under very specific settings. However, exploratory causality can require many interactions between the user and the algorithm until a comprehensive model explaining the observations is achieved. These types of complex analytical processes can be very difficult to manage without interactive visual feedback.

Our proposed visual analytics methodology fills this gap, integrating the human tightly into the temporal causal analysis process. 
We leverage \textit{logic-based causality}, as established by Kleinberg et al. \cite{Kleinberg2009,Kleinberg2011} where a proposition (a.k.a. a phenomenon or an event) in time 
is defined as the time points at which a variable's value falls into a specified range. An event is considered a \textit{potential} cause if it elevates the effect after a certain time delay; it is then scored against other potential causes to test if it is a better explanation. We decomposed this process into several task-based phases 
and then implemented this pipeline into a prototype system, DOMINO.  It supports users in determining relevant causal relations and the time windows at which they occur via simple mouse interactions.
Several complementary visualizations are provided to explore and confirm these relations.


In summary, the major contributions of our paper are:
\begin{itemize}
\item A visual analytics methodology that inserts human users into the loop of temporal causality analysis.
\item An extension of logic-based causality theory to allow for the automated search of potential causes.
\item A prototype visual system that enables users to both generate and test temporal causal hypotheses. 
\item A collection of scenarios and expert feedback demonstrating the capabilities of our causality framework.
\end{itemize}

We note that logic-based causality (and likewise Granger causality) does not produce a causal graph or network. Rather, it looks to explain to what extent the various system variables are able to consistently raise or lower the values of a specific effect variable, yielding a set of causal relations. 

Our paper is structured as follows. Section 2 discusses related work. Section 3 introduces Kleinberg's logic-based causality and our extensions. Section 4 describes design goals and the analytical workflow our system facilitates. Section 5 describes the various components of the visual interface. Section 6 illustrates 3 usage scenarios. Section 7 presents two user studies. Section 8 provides a discussion of our work including ideas on how our method could be generalized to construct entire causal networks. Finally, Section 9 ends with conclusions.

\section{Related Work}

Modern causality theory generally distinguishes two mechanisms of causality -- counterfactual and temporal. With counterfactuals, a causal relation can be verified with a conditional independence test where we test if two variables are correlated under conditional settings \cite{Pearl2000}. Inference algorithms built on such tests (e.g., PC\cite{Spirtes2000}) typically learn an overview of the causal system from all of the derived conditional independence constraints. The inferred graph model can then be parameterized into either Bayesian Networks (BNs) \cite{Friedman2003, Pearl2001} or Structural Equation Models \cite{Pearl2000}.

The temporal definition stresses that a cause always precedes its effect. Theories in this category typically are used for analyzing relations between time series. Some of the most studied approaches include Dynamic Bayesian networks (DBNs)\cite{Murphy2000} and Granger causality\cite{Granger1969}. A DBN extends the causal BN and uses two networks to represent the system at any two consecutive time points. Then, connections are added between nodes of the two time slices to describe how the system evolves over time. This assumes that both the structure and the states of the system will repeat after some initial state. In contrast, Granger causality takes several time series and determines if one can be predicted by another with some time lag. However, since the test is carried out simply with auto-regressions, Granger causality is actually similar to correlation.

Finally, Logic-based causality \cite{Kleinberg2009,Kleinberg2011} was devised more recently for analyzing the dependencies among temporal events. It depicts causality as hypothetical relations between logic propositions with arbitrary time lag. The true causes among all potential causes can then be identified via significance tests. It offers a very general and flexible formulation for extracting causal relationships in time series data. Our work builds on this theory and makes it more accessible to analysts via a carefully designed visual analytics system. 

\subsection{Visual Analytics of Causality}

As stated by Chen et al. \cite{chen2011}, the goal of a visual analytics solution for causality is to support decision making in business settings, scientific investigations, etc. The authors further state that these systems should provide the ability to both formulate and evaluate hypotheses and so stimulate creative thinking. An earlier attempt along these lines is the \textit{Growing-polygons} system \cite{Elmqvist2004} which captures causality as a sequence of causal events (a.k.a., \textit{process}), and uses animated polygon sizes and colors to signify causal semantics. Adopting the idea of causal graphs, Vigueras and Botia \cite{Vigueras2007} consider ordered events in a distributed system as causally dependent and visualize their relations in colored 2D graphs. \textit{ReactionFlow} \cite{Dang2015} arranges duplicate variables in two columns and visualizes causal relations between them as pairwise pathways, assisting user queries along the causal chains.

There are several methods that use causal models to enable what-if analyses in algorithmic decision making. Yan  et  al.  \cite{ yan2020silva} propose  a  method by which analysts can  edit a post-hoc causal model in order to reveal issues with algorithmic fairness. Hoque and Mueller \cite{hoque2021outcome} allow users to interactively vary the values of a model’s exogenous and endogenous variables and observe the outcome.  Xie  et  al. \cite{ xie2020visual} work in a similar direction but focus on probability distributions. None of these methods consider temporal relations.

Taking time delay into consideration, Li et al. \cite{Li2012} used Granger causality to measure the activity of brain neurons and built a 3D visual analytics system for this task. More recently, \textit{DIN-Viz} \cite{forbes2018} was devised as a visual system for analyzing causal interactions between nodes in influence graphs simulated over time. Bae et al. \cite{bae2017} evaluated representations of causal graphs and claim that arrows or tapered edges can result in better recognizability. Although effective in causality visualization, none of the aforementioned works offers an automated inference function, and so all must rely solely on user input for the initial causal relations.

The first visual system with automated causal reasoning was proposed recently by Wang and Mueller \cite{Wang2016}. It utilizes constraint based algorithms and provides a set of interactive tools that allow the user to examine the derived relations. A further development of this work offers the capability of analyzing different models that may inhabit separate data subdivisions. It also improves the causal graph visualizations by expressing them as color-coded flow diagrams \cite{Wang2017}. As the underlying methods do not consider time, the system cannot be used for analyzing temporal dependencies. 

\subsection{Visual Analytics of Multivariate Temporal Data}
For this discussion we shall distinguish among two types of temporal data, namely, \textit{time series data} which are a set of observations obtained through repeated measurements over time, and \textit{event sequence data} which are a set of discrete events ordered in time of occurrence. While the former is typically collected at regular intervals, the latter may not be. Often, event sequence data are derived from raw time series data via feature extraction and pattern analysis. 

Many works on time series analysis have been focusing on efficient algorithms to identify frequent patterns, either in a supervised \cite{pham2009} or unsupervised fashion \cite{rodrigues2008, kiernan2009}. Methods based on deep neural networks \cite{katarya2018}
have gained much attention recently as they promise fully automated solutions for pattern identification and prediction. However, explaining to human analysts the reasoning behind such identifications and predictions remains challenging.

Early work on the visual analysis of event sequence data includes \textit{Lifelines} \cite{plaisant1996} which emphasizes detailed records of a single patient. A follow-up version, \textit{Lifelines2} \cite{wang2009}, utilizes a set of dynamic aggregations to highlight the prevalence of events occurring in multiple sequences. Zgraggen et al. \cite{Zgraggen2015} propose a touch-based system featuring a visual query language for time series built via regular expressions. \textit{WireVis} \cite{Chang2007} connects temporal events by monitoring a set of user-defined keywords and visualizing the detected relations as a network. Lee and Shen \cite{Lee2009} define salient local features in time series as \textit{trends} and utilize visual tools for matching and grouping similar patterns. Zhao et al. \cite{Zhao2017} aggregate large numbers of short sequences into connected matrices such that common patterns can be easily observed.  To investigate the common progressions of events, \textit{DecisionFlow} \cite{Gotz2014} and its related projects \cite{Wongsuphasawat2012, Monroe2013} consolidate temporal events as alternative pathways with embedded patterns. The recent survey by Guo et al.~\cite{guo2021survey} provides a comprehensive overview on the visual analysis of event sequence data. 

Recent work by Jin et al. \cite{jin2020visual} proposes a visual analytics method to aid users in the construction of general causal models from large collections of multivariate event sequence data. Their system uses Granger causality on Hawkes processes to derive a set of causal relations that form the basis of the proposed user-assisted model construction procedure. Our work differs from their's in that we begin with a single raw multivariate time series dataset, and our primary goal is to help analysts understand the causal relations, and their time windows, that govern the specific phenomena captured by these data. As such, our system aids users in the extraction of events under the premise of time-dependent causality, interactively with visual support.


\section{Theoretical Foundations and Algorithm} \label{sec:theory}

In logic-based causality \cite{Kleinberg2009}, a causal hypothesis is a presumed relationship between several logic propositions with a time lag. A proposition describes an observed phenomenon or event, such as wind speed $<$ 15 km/h, or blood glucose level of 70-100 mg/dl which is the normal blood sugar level before a meal for a human without diabetes. Mathematically, this is modeled by a \textit{state formula} testing if a variable satisfies a numerical constraint, e.g., $a\le 4.1$ or $b\in [10,18]$ where $a$ and $b$ are observed variables.

Given two state formulas $c$ and $e$ where $c$ causes $e$, a \textit{path formula} specifies the direction, the strength, and the window of time delay of the causal relation. This is formally written in \textit{leads-to} notation as
$c\leadsto ^{\ge r, \le s}_{\ge p}e$
which means if $c$ is true $e$ will become true with a probability of at least $p$ after a delay between $r$ and $s$ time units, where $0\le r\le s \le\infty$. For example, in the hypothesis of ``smoking causes cancer in 5 to 10 years with 55\% probability", the propositions of [smoking=True] and [cancer=True] each gives rise to a state formula, and the path formula hypothesizes a 55\% chance that the causal relation will happen after 5 to 10 years.

Let $T$ be a time sequence. A time point $t$ in $T$ satisfying a state formula $c$ is written as $t\models_T c$ and a subsequence of time points $\pi^t$ starting from $t$ satisfying path formula $c\leadsto ^{\ge r, \le s}e$ is written as $\pi^t \models_T c\leadsto ^{\ge r, \le s}e$. Then the probability of the path formula is calculated as
\begin{equation} \label{eq:path-formula-1}
P(c\leadsto ^{\ge r, \le s}e)=\frac{|\{t\in T: \pi^t \models_T c\leadsto ^{\ge r, \le s}e\}|}{|\{t\in T: t\models_T c\}|}
\end{equation}
which is the number of times the relation holds divided by the number of times the cause event occurs.

Although a state formula in classic logic-based causality theory can be made of multiple propositions and can be defined recursively, we assume for the remainder of this paper (for convenience) that there is only one atomic proposition in each state formula, and one proposition corresponds to one event/phenomenon. To check the truth values of a conjunction of multiple state formulas, we can simply check the label of each event at every time point, and then merge the labels at a matching time with a logical $and$ operation.

\begin{figure}[!t]
 \centering
 \includegraphics[width=0.8\columnwidth]{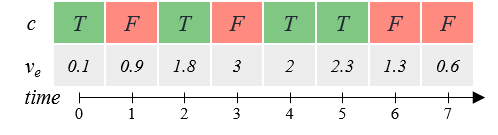}
 \caption{Inferring a potential cause. Assuming we have a short sequence of a continuous variable $v_e$ and an event $c$. Averaging all values of $v_e$, we have $E[v_e]$=$1.5$. With a time delay of 1 unit, $E[v_e|c]$ = $(0.9+3+2.3+1.3)/4$=$1.875$, so $c$ should be a potential cause of $v_e$ according to Eq.~\ref{eq:prima-exp}. If we instead make an event $e$=$[v_e>E[e]]$, we will have $P(e)$=$0.5$ and $P(e|c)$=$0.5$, such that $c$ is not a potential cause of $e$.}
 \vspace{-10pt}
 \label{fig:potential-cause}
\end{figure}

\subsection{Inferring Causes}

The testing of an event $c$ being a cause of the effect $e$ is based on the assumption that a true cause always increases the probability of the effect (we can view a preventative, which lowers the probability of $e$, as raising the probability of $\neg e$). Thus, we say $c$ is a \textit{potential cause} \cite{Kleinberg2009} of $e$ if, taking into consideration the relative time delay window, it satisfies:
\begin{equation} \label{eq:prima-prob}
P(e) < p \textrm{ and } P(e|c) \ge p
\end{equation}
where $P(e|c)$ is calculated in the same fashion as Eq.~\ref{eq:path-formula-1}.

Additionally, if the effect $e$ is defined on a continuous variable $v_e$ and we are looking for events that are potentially shifting the distribution of $v_e$ (as opposed to a value of $v_e$ falling into a specific range), the expected value of $v_e$ can be used instead for better sensitivity to change \cite{Kleinberg2011}. As such, $c$ is considered a potential cause of $e$ when
\begin{equation} \label{eq:prima-exp}
E[v_e]\neq E[v_e|c]
\end{equation}
Here, the $\neq$ sign can be replaced by either $>$ or $<$ to stipulate only positive or negative causes. And the conditional expected value can be calculated as
\begin{equation} \label{eq:prima-exp-1}
E[v_e|c]=\sum_y y\frac{\Theta([v_e=y]\wedge c)}{\Theta(c)}
\end{equation}
where $y$ are values in $v_e$'s domain and $\Theta(x)$ denotes the number of time points where $x$ holds.

To illustrate, Fig.~\ref{fig:potential-cause} shows short sequences of a continuous variable $v_e$ and a causal event $c$. Averaging all values of $v_e$, we have $E[v_e]$=$1.5$. Then, when considering a time shift of exactly 1 unit, we have $E[v_e|c]=(0.9+3+2.3+1.3)/4=1.875$ (these are the values of $v_e$ exactly 1 unit after $c$ is $T$). Since $E[v_e|c] > E[e]$, according to Eq.~\ref{eq:prima-exp}, $c$ increases the expected value of $v_e$ and thus is a potential cause of it. 

However, if we try to find the positive cause by instead bounding $v_e$ to a specific range, or to a specific value such as the mean of $v_e$ the event $e$ would be defined as [$v_e > E[v_e]$]. Then we would have $P(e)=0.5$ ($e$ occurs 4 times out of 8 time points) and $P(e|c)=0.5$ (2 out of 4), where $c$ would not be considered a potential cause because it is not raising the probability of $e$. This shows the reduced sensitivity to change that comes with trying to be more specific. 

We can generalize this framework to a set of causes $X$ of an effect $e$. We measure the influence of $X$ towards $e$ by calculating the probability change of $e$ as $P(e|X)-P(e)$ or the change of expected value of $v_e$ as $E[v_e|X]-E[v_e]$, depending on the definition of $e$. Note that while the conditional probability is bounded within $[0, 1]$, the expected value could be any amount, and either positive or negative.


\begin{figure}[!t]
\vspace{-10pt}
	\centering
	\subfigure[]{\includegraphics[width=0.4\linewidth]{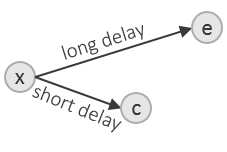}}
	\subfigure[]{\includegraphics[width=0.4\linewidth]{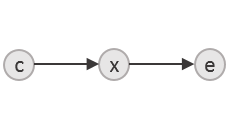}}
	   \vspace{-5pt}
    \caption{The two situations where an event $c$ can be erroneously considered as causing the event $e$ with Eq. \ref{eq:prima-prob} and \ref{eq:prima-exp}. (a) $c$ and $e$ are independent but are commonly caused by the confounder event $x$ with $c$ being caused earlier than $e$. (b) $c$ causes $e$ by mediation via another event $x$.}
    \label{fig:err}
    \vspace{-10pt}
\end{figure}

\subsection{Dealing With Confounders and Mediators}

 As mentioned, a causal relation is only potential; it may not be direct or it may be spurious, even if Eq.~\ref{eq:prima-prob} or \ref{eq:prima-exp} holds. This can be due to two possible situations: (1) $c$ and $e$ are actually independent but are commonly caused by another event $x$ (the \textit{confounder}) with $c$ being caused earlier than $e$ (Fig.~\ref{fig:err}a), or (2) $c$ causes $e$ indirectly via $x$ (\textit{mediation}, Fig.~\ref{fig:err}b). In either condition, we may observe that Eq.~\ref{eq:prima-prob} or \ref{eq:prima-exp} holds and erroneously mark $c$ as directly causing $e$. One way to eliminate such error is to compare the distribution of $e$ when $c$ and $x$ both occur, i.e., $P(e|c\wedge x)$, to that when only $x$ is present, i.e., $P(e|\neg c\wedge x)$. Then the two will be found equal (or almost equal) if $c$ is a spurious cause of $e$. Note that this requires the time window $[r,s]$ to be sufficiently wide such that both $x$ and $c$ could have caused $e$ \cite{Kleinberg2010}.  

This idea can be generalized to multivariate time series. When considering multiple time series in a dataset, for a given effect, we usually can recognize a number of potential causes. To identify the real causes that can better explain the effect, Eells \cite{Eells1991} proposed the average significance of a potential cause $c$, among all potential causes $X$ toward the effect $e$, calculated as
\begin{equation} \label{eq:sig-prob}
\varepsilon_{avg}(c, e) = \sum_{x\in X/ c} \frac{P(e|c\wedge x) - P(e|\neg c\wedge x)}{|X/ c|}
\end{equation}
Here $X/ c$ is the set of potential causes excluding $c$ and $|X/ c|$ is the number of events in it. We need at least two potential causes to make the computation meaningful and all calculations are associated with a preset time window. Then, by setting a certain threshold $\varepsilon$, $c$ is called an $\varepsilon$-significant cause of $e$ if $|\varepsilon_{avg} (c,e)|\ge\varepsilon$. Further, if $e$ stands for the increase or decrease of a continuous variable $v_e$ over the time window, the conditional probability in Eq.~\ref{eq:sig-prob} can be replaced by the conditional expected value such that
\begin{equation} \label{eq:sig-exp}
\varepsilon_{avg}(c, e) = \sum_{x\in X/ c} \frac{E[v_e|c\wedge x] - E[v_e|\neg c\wedge x]}{|X/ c|}
\end{equation}

Although the $\varepsilon$ threshold is decisive in testing if a cause is significant, its value can be difficult to determine automatically in practice. In presence of a large number of (say, thousands of) potential causes where significant causes are rare, all such $\varepsilon_{avg}$ values usually follow a Gaussian distribution \cite{Kleinberg2011}. As a result, the problem can be solved by testing the significance of a null hypothesis where $p$ values rejecting the null hypothesis deviate from the mean \cite{Efron2004}. However, we find that this theoretical method cannot really be applied in most of our applications since we rarely encounter such a large number of time-series and causal events, especially when we just wish to explore the impact of some specific causes on the target. In such cases, the $\varepsilon$ threshold can only be assigned empirically and interactively by the analyst. This requirement for user assistance, together with other analytical tasks that will be discussed later, motivated the visual analytics system that is at the heart of our work.

\begin{figure*}[!tb]
\centering
\includegraphics[width=0.9\linewidth]{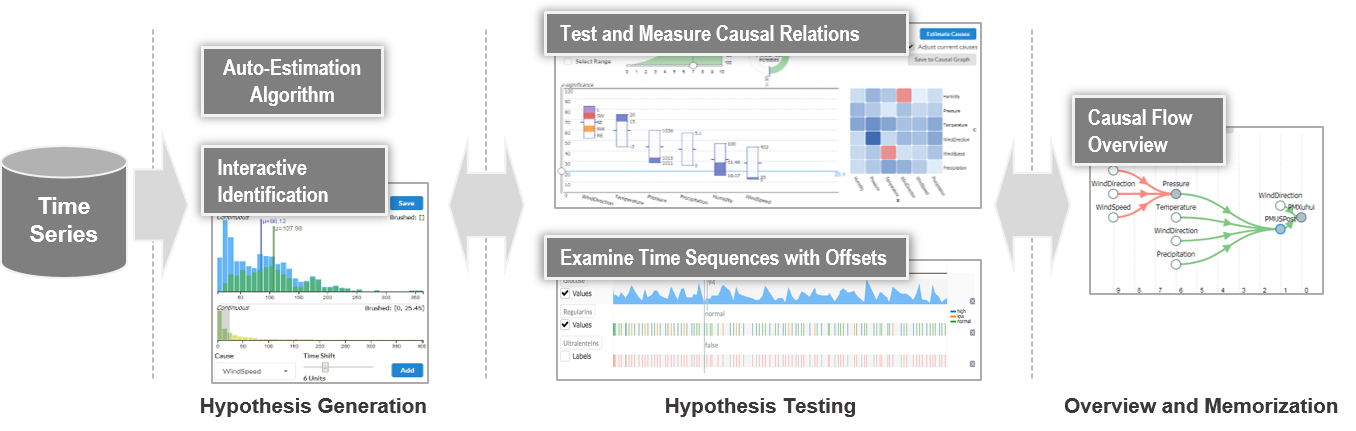}
\vspace{-10pt}
\caption{The analytical workflow of temporal causality analysis. The user first generates hypotheses either interactively or automatically. The added events are then tested and visualized. After iteratively adjusting the time delay and event constraints, the final results can be saved and revisited.}
\label{fig:pipeline}
\vspace{-10pt}
\end{figure*}

\subsection{Automated Estimation of Potential Causes} \label{subsec:auto-est}

Since a potential cause elevates the probability (Eq.~\ref{eq:prima-prob}) or alters the expected value (Eq.~\ref{eq:prima-exp}) of the effect, the process of searching for a cause $c$ is the same as deciding an appropriate numerical constraint on the cause variable $v_c$, on which $c$ is made, so that Eq.~\ref{eq:prima-prob} or \ref{eq:prima-exp} can be satisfied. This is relatively easy and straightforward when $v_c$ has discrete values, where we can simply scan through $v_c$'s domain and make $c$ take all the values satisfying the condition.

The search becomes more complex when $v_c$'s domain is continuous. One solution could be to discretize $v_c$ and then apply the same scanning process, but determining a good discretization strategy is difficult. Our approach is to instead only look at $v_c$ at time points $t$, noted as $v_c(t)$, where $e$ holds after the specified time delay ($v_c(t)\leadsto^{\ge r, \le s} e$). Recording all such $v_c(t)$ as $T_c$, we then discretize $v_c$ adaptively by clustering values in $T_c$. The idea is to consider values that $v_c$ frequently takes and lead to $e$ as possibly causing $e$. 

\begin{algorithm} [!b]\label{alg:1}
    \KwIn{The effect event $e$, a continuous variable $v_c$, distance threshold $\theta$, max iteration $n$}
    \KwOut{A potential cause $c$ of $e$}
    
    $T_c=\{v_c(t) | v_c(t) \leadsto ^{\ge r, \le s}e\} \}$\;
    $i=1$\
    $clusters=$[\{the first value in $T_c$\}]\;
    \While{$i \le n$ and clusters not converging}{
    	\ForEach{$v \in T_c$}{
        	\ForEach{cluster center $\alpha \in$ clusters}{
            	\eIf {$|\alpha - v| < \theta$} { Add $v$ to $\alpha$\; }
            	{ Make $v$ a new center in $clusters$\; }
            }
        }
        Update cluster centers and remove outliers\;
        $i = i + 1$\;
    }
    \ForEach{$k\in clusters$}{
    	Make $c_k$ an event on $v_c$ with value range of $k$\;
        Check if $c_k$ satisfies Eq. \ref{eq:prima-prob} or \ref{eq:prima-exp}\;
    }
    Merge events if their value constraints overlap\;
    \Return the event $c\in\{c_k\}$ that maximally elevates $e$
    \caption{Estimating a potential cause}
\end{algorithm}

The clustering process follows Algorithm \ref{alg:1}, which takes the same approach as the incremental clustering for high-dimensional data \cite{Wang2017a} but is applied in 1-D. The original algorithm has shown great performance in learning informative value segments. The adapted version here iteratively scans values in $T_c$ until  all clusters converge or the maximum number of iterations is reached. In each iteration, a value is assigned to a cluster center if the distance between them is smaller than some threshold $\theta$. A new cluster is added when a point is too far away from all clusters. The threshold $\theta$ controls the size of the clusters, which decides how $v_c$ will be discretized later. At last, we transform $v_c$ by considering the value range each cluster covers as a level, and test if it fulfills Eq. \ref{eq:prima-prob} or \ref{eq:prima-exp}. If multiple levels are returned, we try to merge them if they overlap and take the one that best elevates $e$ as the most possible cause.

The algorithm can be easily parallelized \cite{Wang2017a} if we modify the incremental process such that it searches clusters in batches instead of inspecting them one by one, enabling scalability. Also, the trade-off of taking different $\theta$ values is that a larger $\theta$ tends to produce a looser constraint (a larger value range of $v_c$) in $c$, often resulting in a smaller $P(e|c)$ or an $E[v_e|c]$ closer to $E[v_e]$ -- a smaller $\theta$ does the opposite. This is similar to the problem of under-/over-fitting. In our experiments, we found when $\theta$ equals 0.15 of $v_c$'s value range 5 iterations were usually sufficient to reach a plausible result -- one that would be close to manual adjustment.

\section{Design Goals and Workflow}{\label{sec:task}}

We began by conducting a comprehensive literature review on causality theories, especially logic-based causality, and their applications. Based on this study, we identified 4 high-level analytical tasks and a 4-phase practical workflow to guide the development of our interactive visual interface.

\medskip
\hangindent=2em
\hangafter=1
\noindent
\textbf{T1. Generating causal propositions and hypotheses} is often the first step in causality analysis. Most current works on temporal causality achieve this either by manually grouping relevant values and then assigning them semantic meanings \cite{stanton2015,Zhang2015,Prabhakar2010}, or by conducting an exhaustive search after evenly partitioning the time series data into a large amount of sections each considered an event \cite{Kleinberg2010,Kleinberg2011}. Both of these approaches are limited in efficiency and flexibility. Since in logic-based causality a causal relation is defined over a time lagged conditional distribution \cite{Kleinberg2010}, analysts should be given direct access to such information by allowing them to generate causal propositions and hypotheses with visual support. Also, since an effect can have multiple causes, an overview of the values and boolean labels of each time series in a synchronized fashion will help the understanding of the compound relations.

\smallskip
\hangindent=2em
\hangafter=1
\noindent
\textbf{T2. Identifying significant causes under a specified time delay} is the most common task when investigating causality within time series. Examples are found in temporal causal analysis of the stock market \cite{Kleinberg2010}, biomedical data \cite{Heintzman2016,Hripcsak2011,Li2012}, social activities \cite{Bui2016}, and terrorist activities \cite{stanton2015}. While the significance threshold determining the truthfulness of causes may often need to be decided empirically, a visual system should effectively externalize the identified causes and their levels of significance under the specified time delay, supporting the analyst's decision-making process.

\smallskip
\hangindent=2em
\hangafter=1
\noindent
\textbf{T3. Analyzing the change of causal influences over time} is another important task as the significance and influence of a cause toward the effect could differ over settings of time delay. Interactive operations can reveal the time span of a causal relation, as well as a proper window of time delay for identifying other causes. The latter, however, has been mostly assigned empirically with a limited set of values in the related work so far \cite{Kleinberg2010, Heintzman2016, Li2012}. When this knowledge is incomplete, a visual system should support analysts in these tasks by visualizing the causal influences toward the effect associated with all possible time delays under consideration.

\smallskip
\hangindent=2em
\hangafter=1
\noindent
\textbf{T4. Interactive analysis.} As mentioned, causality analysis is often associated with a number of meta parameters to be determined by analysts empirically, e.g., the numerical constraints in the causal propositions and the threshold in the significance tests. Determining these parameters is an essential task in causality analysis and often requires interaction, as illustrated in other projects \cite{Wang2017,Wang2016,Dang2015}. To support interactive analysis, an effective system should provide visual feedback along with each of the user's parameter updates. Users should also be able to save the discoveries in an overview for later re-examination.

\begin{figure*}[!tb]
  \centering
  \includegraphics[width=0.9\linewidth]{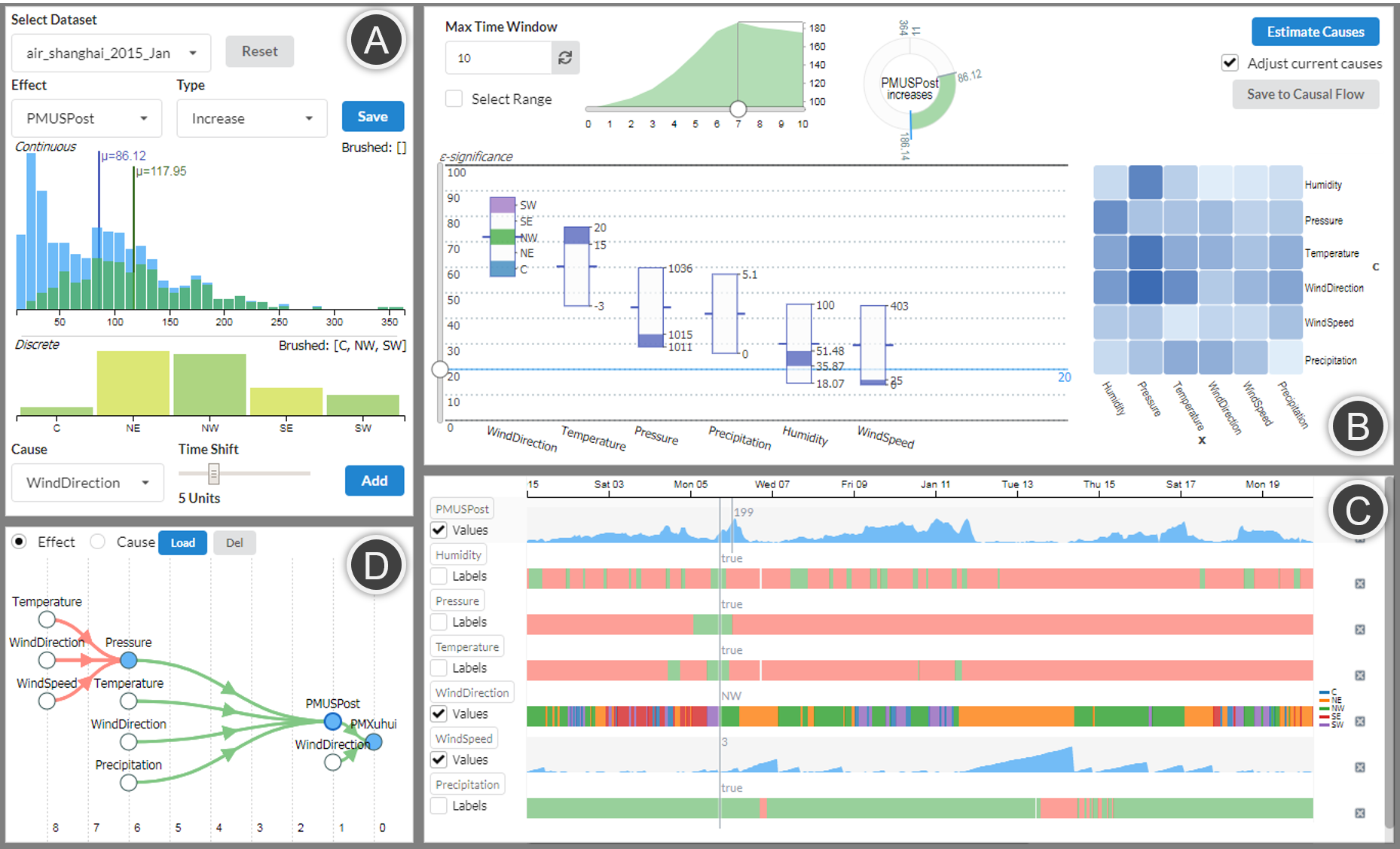}
  \caption{The DOMINO interface analyzing the Air Quality dataset. The interface consists of (A) the conditional distribution view for manually exploring potential causes of the specified effect, (B) the causal inference panel for the interactive analysis of causal relations under different time delays and significance thresholds, (C) the time sequence view for examining the synchronized time series, and (D) the causal flow chart displaying the established causal relations. The interface is designed to follow the analytical process of the temporal causality analysis established in Section ~\ref{sec:task}.}
  \vspace{-5pt}
  \label{fig:overview}
\end{figure*}

\smallskip
To guide the design of a dedicated visual system that can support temporal causality analysis as an interactive process of generating and testing causal hypotheses under time windows, we devised an analytical workflow comprising the following four phases (see Fig.~\ref{fig:pipeline}):

\smallskip
\hangindent=2em
\hangafter=1
\noindent
\textbf{P1. Causal hypothesis generation} is the first step of the pipeline where a user specifies the target effect and then locates its hypothetical causes (\textit{T1}). This can be done either in an interactive approach (\textit{T4}) or with auto estimation algorithms (e.g., Algorithm \ref{alg:1}). Either way, the window of time delay is adjustable (\textit{T3}).

\smallskip
\hangindent=2em
\hangafter=1
\noindent
\textbf{P2. Testing the hypothesis} under its statistical significance is the next step after identifying a set of potential causes (\textit{T2}). Besides measuring the inferred causal relations under different thresholds (\textit{T4}), users will also visually examine the time sequences looking for direct evidences under different settings of delay (\textit{T3}).

\smallskip
\hangindent=2em
\hangafter=1
\noindent
\textbf{P3. Iterative analysis} is an often required step where a user shifts between adjusting the time delay to evaluate the change of impact of current causes on the effect (\textit{T3}), and adjusting the causes regarding their constraints looking for alternatives under the current setting (\textit{T2}).

\smallskip
\hangindent=2em
\hangafter=1
\noindent
\textbf{P4. Overview and memorization} is typically the final step where a user saves his/her discoveries for later access or integrates them into an evolving causal network (\textit{T4}).


\section{The DOMINO Implementation} \label{sec:impl}

\noindent The DOMINO system we present in the following fully implements the analytical pipeline introduced in Section \ref{sec:task}. An overview of the interface is shown in Fig. \ref{fig:overview}, analyzing the Air Quality dataset (see Section \ref{sec:case} for details). The interface is composed of four linked panels, A-D, explained next.

\subsection{The Conditional Distribution View (Panel A)}
Our design of the conditional distribution view was motivated by the definition of a potential cause mentioned in Section ~\ref{sec:theory}. With the conditional distributions displayed here, analysts can directly observe a time-lagged phenomenon and hence make causal hypotheses (\textit{T1,T4}). 

Upon selecting and loading a tabular data file of a multivariate time series (top menu) the analyst selects one of the variables as the effect variable \textit{e} and specifies the effect of interest, such as \textit{Value Increase}, \textit{Value Decrease}, or \textit{Value Range} (second menu row). Upon selecting \textit{e} the system displays a histogram of its values $v_e$ over the entire time series duration (blue bars). This histogram can be brushed when the selected effect is \textit{Value Range} (the \textit{Brushed} box, top right, must be checked). Next, the analyst selects a cause variable in the menu on the bottom left, and a frequency histogram of the variable's levels (if discrete) or values (if continuous) will be shown above the menu. The analyst then indicates the potential cause \textit{c} by selecting the corresponding level bar(s) and specifying a time delay via the slider on the right. Following, the system displays a histogram of \textit{e} when \textit{c}'s event conditions are met (green bars). This histogram is necessarily lower in magnitude since the conditions are only met for some $v_e$. The vertical bars for the histograms indicate their respective means, $E[v_e]$ and $E[v_e|c]$. The wider apart these means the more pronounced \textit{c}'s potential effect.  

\begin{figure}[!t]
	\centering
  	\subfigure[]{\includegraphics[height=0.82in]{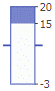}}
  	\subfigure[]{\includegraphics[height=0.78in]{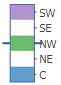}}
	\subfigure[]{\includegraphics[height=0.78in]{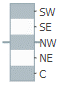}}
    \subfigure[]{\includegraphics[height=.93in]{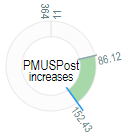}}
  	\subfigure[]{\includegraphics[height=.93in]{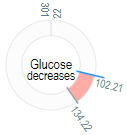}}%
  	\caption{Visual encoding of events in the causal inference panel. (a) A box in the box chart representing a significant cause exerted by a continuous variable; only the significant value range is colored. (b) A significant cause exerted by a discrete variable; only the significant levels are colored. (c) An insignificant cause. (d) A positive effect (\textit{Increase} type) with elevated expected value. (e) A negative effect (\textit{Decrease} type). The colors in (b), defined in a D3 colormap, are used for the same levels also in the discrete variable's time sequence view (see Fig. 1C and Fig. 9(b)).}
  	\label{fig:event-anno}
  	\vspace{-10pt}
\end{figure}

\subsection{The Causal Inference Panel (Panel B)}
Each time the \textit{Add} button in panel A is pressed, the cause \textit{c} specified there is added to the set of potential causes \textit{X}. DOMINO then automatically tests its significance with regards to \textit{e} and time delay (\textit{T2,T4)}, and positions it as a box in the box chart (bottom left). The boxes are identical in size and ordered by significance; the value is signified by the two small handles attached at the left/right box center. Users can move a vertical slider up/down to adjust the $\varepsilon$-threshold. If there are too many boxes, a horizontal scrollbar will appear for scalability. Fig. \ref{fig:event-anno}a-c gives the visual encoding of the boxes. The colored segments represent the value constraints specified in panel A, annotated by the labels on the right. For a significant cause, its color scheme is decided by its continuous (Fig. \ref{fig:event-anno}a) or discrete (Fig. \ref{fig:event-anno}b) value type, else it is in gray (Fig. \ref{fig:event-anno}c). 

The area chart above the box chart visualizes the combined effect of all significant causes for a range of time delays (\textit{T3}), starting at 0 all the way to the maximum delay selected in the value box on the left. The color encodes the type of effect -- red for \textit{Decrease} and green for \textit{Increase} and \textit{Range}. The slider on the chart's time-axis sets the time delay used in the significance tests. The default is a narrow window that just includes one sample point of the time series. A wider window can be chosen by checking the \textit{Select Range} box on the left. This will expose two handles by which the window width $[r,s]$ can be specified. 

The area chart is fully integrated with the other interface components. Whenever the user moves the slider or handles, all other charts update accordingly. Likewise, moving the threshold slider in the box chart will also update the area chart to reflect the change in the selection of significant causes (\textit{T4}). The donut chart next to the area chart uses the same color scheme than the area chart and shows the effect variable's expected value (or probability change) at the set time delay and window at more precise detail. Here, the grey and blue colored indicators reflect $E[v_e]$ (or $P(e)$) and $E[v_e|c]$ (or $P(e|c)$), respectively (see Fig. \ref{fig:event-anno}d, e). 

The donut chart is essentially a vertical box warped into a ring where the bottom and top extreme values meet at the 12 o’clock position. We chose this representation to distinguish it from the vertical boxes we used for the cause variables which have a different color mapping. It is also a compact representation that allows the name and trend of the variable to be set as a text string into the chart center.

The matrix chart on the right visualizes the intermediate results from the inference process. Each row and column of the matrix corresponds to a cause. A tile in the diagonal is colored according to the value of $P(e|c) - P(e)$ or $E[v_e|c] - E[v_e]$ in Eq. \ref{eq:prima-prob} and \ref{eq:prima-exp}. A non-diagonal tile at row of cause $c$ and column of cause $x$ is colored by the value of $P(e|c\wedge x)-P(e|\neg c\wedge x)$ or $E[v_e|c\wedge x]-E[v_e|\neg c\wedge x]$ per Eq. \ref{eq:sig-prob} and \ref{eq:sig-exp}. If the effect type is \textit{Decrease}, tiles with negative values will be colored blue and positive red. For \textit{Increase} the opposite scheme applies. The value, and the equation used for computing the value, will pop up as a tooltip when the mouse hovers over a tile. In this way the user can inspect a row to explore its associated cause and then choose a column to check how its impact measures up with the column variable's impact. Hence. the darker the cell the stronger the row variable's impact relative to the column variable's impact.

Finally, panel B also contains the button (upper right corner) by which the automated estimation of potential causes can be initiated, based on the effect variable and type specified in panel A. Execution of this routine will then auto-populate all charts. 

\subsection{The Time Sequence View (Panel C)}
The time sequence view presents an enhanced rendering for examining the time sequences directly (\textit{T1}). The sequences are stacked, with the specified effect on the top. Each sequence can be rendered in two modes -- label mode and value mode, which can be switched by clicking the check box on the left (unchecked is 'labels' and checked is 'values'). The former visualizes the Boolean labels of an event at each time as a strip of colored bars (green for \textit{true}, red for \textit{false}). The value mode displays an area chart if the variable is continuous or a strip of bars colored by the level the discrete variable takes at each time with the legend on the right. The same colors are also used in the box chart in the causal inference panel. In both modes, missing values are left blank and long sequences are scrollable.

A user can click on the variable name of a sequence to revisit and adjust the event's value constraint in the conditional distribution view (\textit{T4}). An event can be removed with the delete button on the right of the sequence. Two indicator lines will be rendered and move along with the mouse pointer. The longer line shows the value or label, depending on the visualization mode, of each cause at the time point hovered on. The other line shows the value of the effect ahead, with the time shift set in the causal inference panel area chart. 

\subsection{The Causal Flow Chart (Panel D)}
After a proper set of causes and the time delay is obtained, the user can save the results into the causal flow chart by clicking the \textit{Save to Causal Flow} button at the top-right corner of the causal inference panel. If some previous results exist, DOMINO will merge them by matching the nodes representing the same event and build a tree structure with the significant causal relations. The causal tree is laid out along a time-axis (with dashed indicator lines) in the fashion of a flow diagram where the distance between a cause and an effect signifies their time lag. For delays with a window, the max time lag is used. A link's color indicates the type of the effect -- red links point to \textit{Decrease} and green to \textit{Increase} or \textit{ValueIn}. The nodes in the chart can be reloaded either as a cause or an effect. DOMINO will automatically decide if it should be reloaded or merged into the current relations. A selected node can also be deleted with the \textit{Del} button.




\begin{figure}[!tb]
 \centering
 \includegraphics[width=2.5in]{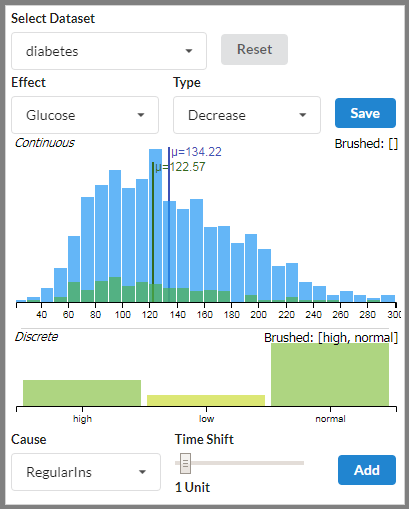}
 \caption{Top view: the conditional distribution view displays the distribution (blue bars) and the conditional distribution (green bars) of the variable \textit{Glucose}. The latter is conditioned on [\textit{RegularIns} $\in\{normal,high\}$] (bottom view) at 1 unit time delay, both selected by the user. The blue (green) vertical lines in the top view show the expected value of \textit{Glucose} before (after) the conditioning. We observe that the green vertical line has a lower expected \textit{Glucose} value, indicating that the selected \textit{RegularIns} dose had the desirable effect after 1 time unit. The user can now interactively select other time shifts and \textit{RegularIns} doses to see the expected \textit{Glucose} value responses to these settings.}
  \vspace{-5pt}
 \label{fig:glu-cdv}
\end{figure}

\begin{figure}[!t]
 \centering
 \includegraphics[width=\linewidth]{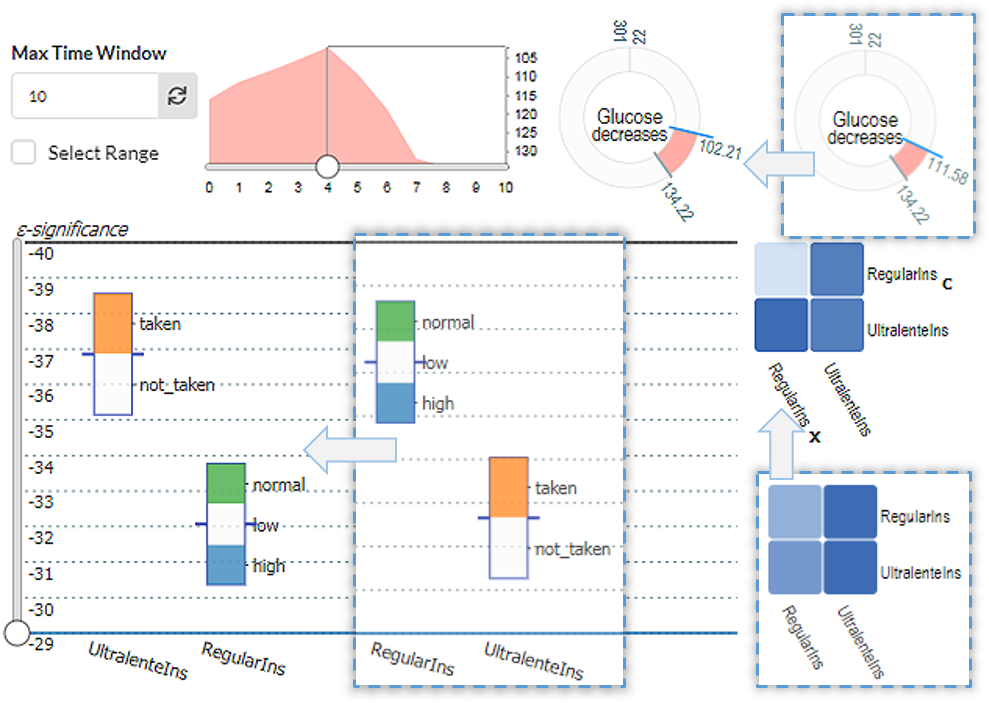}
 \caption{The causal inference panel analyzing the influence of \textit{RegularIns} and \textit{UltralenteIns} on \textit{Glucose}. The three dashed, superimposed visualizations with arrows show the box, donut, and matrix charts after a 1 unit time delay, while the original charts show the values after a 4 unit delay. We observe that after 1 time unit \textit{RegularIns} is a more effective cause to lower \textit{Glucose}. However, with a 4 unit delay (as selected in the area chart), \textit{UltralenteIns} becomes the more significant cause. In the area chart the y-axis is always set to the currently selected effect, here \textit{Glucose}. It has the (more desirable) lower values at the top of the scale since \textit{Decrease} was selected in the Distribution View. The area chart also shows that the minimum \textit{Glucose} (the peak of the area) is at a time delay of 4 units, and so, after moving the area chart slider to this position the dominant effect of \textit{UltralenteIns} is revealed. However, \textit{RegularIns} also plays a supportive role as can be seen in the box chart.   }
 \label{fig:ins}
\vspace{+3pt}
\end{figure}

\begin{figure*}[!t]
 \centering
 \includegraphics[width=6.5in]{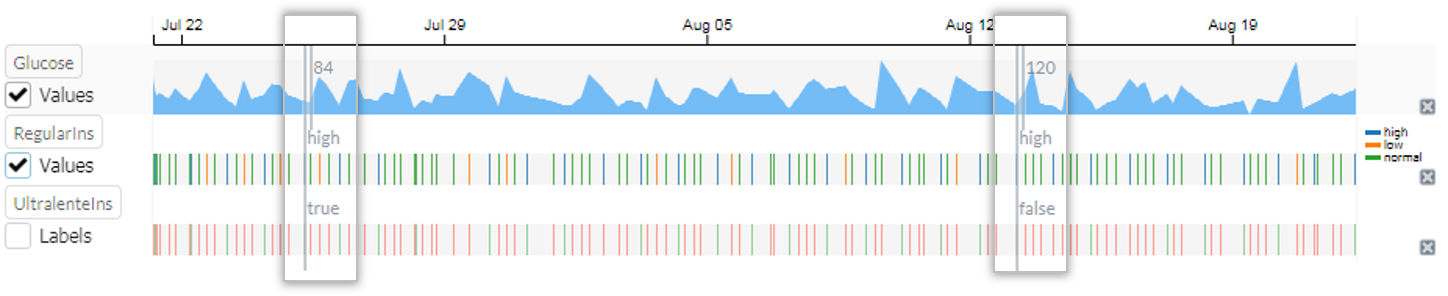}
 \vspace{-10pt}
 \caption{The time sequence view visualizing the glucose dataset under 4 units time offset. The highlighted areas are from two screenshots when moving the mouse pointer over the sequences. The left inset shows that when the \textit{RegularIns} dose is \textit{high} and \textit{UntralenteIns} is \textit{taken} then the \textit{Glucose} will reduce to 84 after 4 time units (short vertical line) after the time point hovered on by the mouse (long vertical line). Conversely, the right inset shows what happens when \textit{UntralenteIns} is \textit{not taken}. In that case the \textit{Glucose} level remains at 120 after 4 time units. Thus, we learn that \textit{RegularIns} is best taken together with \textit{UntralenteIns} to reduce \textit{Glucose} levels.}
 \label{fig:ins-ts}
\end{figure*}

\begin{figure*} [t]
\centering
\begin{tabular} {cc}
    \setlength{\tabcolsep}{0.5pt}
    \renewcommand{\arraystretch}{0.5}
        \begin{tabular} {c}
        \includegraphics[width=4.95in]{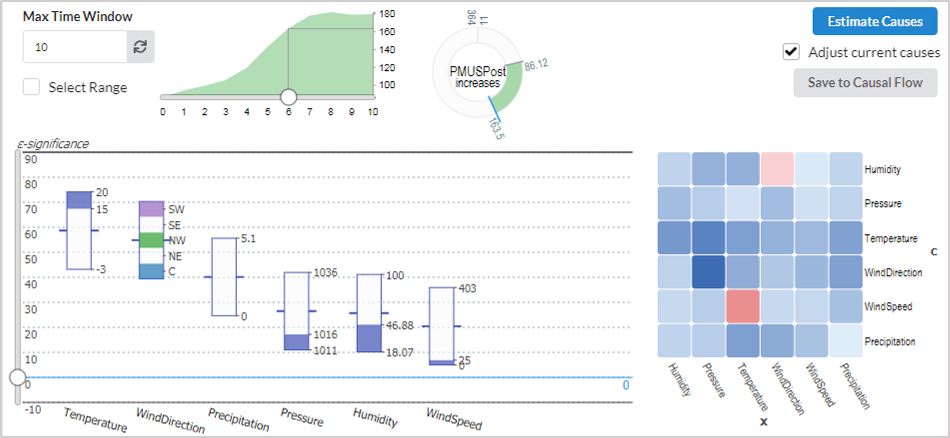}\\
        (a)\\
        \includegraphics[width=4.95in]{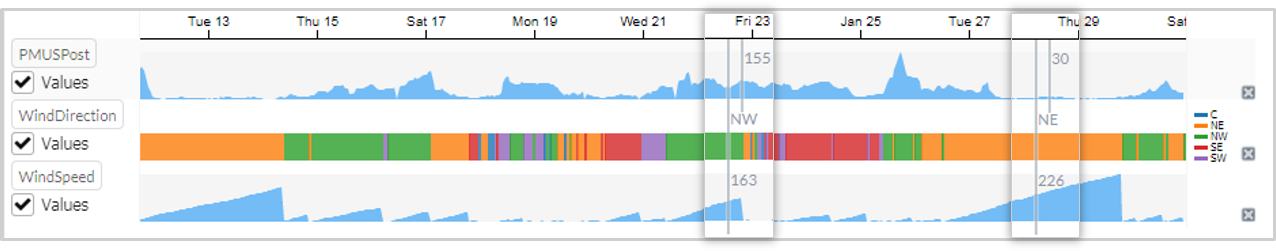}\\
        (b)
        \end{tabular} 
        &
        \begin{tabular} {c}
        \includegraphics[width=1.6in]{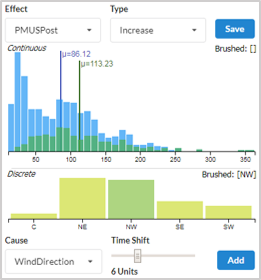}\\
        (c)\\
        \includegraphics[width=1.6in]{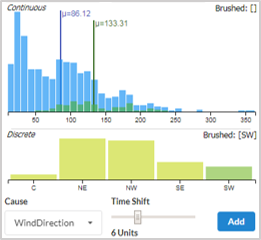}\\
        (d)
        \end{tabular} 
\end{tabular}
\vspace{-10pt}
\caption{Analyzing the Air Quality dataset. (a) The causes increasing \textit{PMUSPost} are estimated automatically by DOMINO, time delay set to 6 hours. (b) Examining the time sequence view reveals that, while wind from the northeast reduces air pollution, wind from the northwest does not. Further analysis comparing (c) the influence of northwest wind and (d) of southwest wind on \textit{PMUSPost} implies that the latter is the larger pollution source.}
\label{fig:case-air}
\vspace{-10pt}
\end{figure*}


\section{Usage Scenarios}{\label{sec:case}}

In this section we demonstrate three usage scenarios featuring DOMINO, using the following three real-world datasets.

\smallskip
\noindent
\textbf{The Glucose Dataset} is part of a complex dataset \cite{Dua2017}, with three time series recorded in a 1-hour interval, monitoring a patient's intake of two types of insulin (\textit{RegularIns} and \textit{UltralenteIns}), and the blood glucose level (\textit{Glucose}). The patient took \textit{RegularIns} regularly at a \textit{low}, \textit{normal}, or \textit{high} dose and sometimes took \textit{UltralenteIns} together. We want to see how they reduce the patient's blood glucose differently.

\smallskip
\noindent
\textbf{The Air Quality Dataset} has 8 attributes, each formatted as a sequence of hourly measured PM2.5 concentrations in air and weather conditions in the city of Shanghai, China. The PM2.5 are fine particles with a diameter of about 2.5 $\mu$m and one of the main air pollutants. The data were collected from two locations - the Shanghai US embassy (\textit{PMUSPost}) and the Xuhui district (\textit{PMXuhui}). The variables associated with weather conditions include \textit{Humidity}, \textit{Pressure}, \textit{Temperature}, \textit{WindDirection}, \textit{WindSpeed}, and \textit{Precipitation}. The dataset was retrieved from Kaggle \cite{AQ_kaggle} and spans 5 years. For simplicity, we only use the data of January 2015 (744 time points in total), as it was the worst month in 2015 for Shanghai with respect to average air quality. We selected this dataset to show DOMINO's use for analyzing more complex data.

\smallskip
\noindent
\textbf{The DJIA 30 Dataset}, fetched from the Investors Exchange data service (\url{https://iextrading.com}), reports the daily highest share price of 30 \textit{Dow Jones Industrial Average} (DJIA) companies from 2013 to 2017 (1203 opening days). We use this dataset to show DOMINO's use to support strategizing in financial analysis.

\subsection{Exploratory Causality Analysis: Glucose Dataset}
\noindent Let's assume a physician, Lisa, who wishes to understand how a diabetic patient, Matt, reacts to two types of insulin, Regular and Ultralente. To gather the data, Matt uses an external blood glucose monitor to record his glucose levels at 1-hour intervals while taking the two types of insulin as prescribed by Lisa. After collecting the data, Lisa decides to use DOMINO to examine the effects of these treatments on Matt's glucose levels over time.   

Lisa first seeks to manually explore these effects. She obtains the conditional distribution view shown in Fig.~\ref{fig:glu-cdv} with \textit{Glucose} selected as the effect variable. The blue histogram displays the distribution of all obtained glucose measurements with a vertical blue line indicating a mean value of $\mu=134.22$. Lisa then selects the effect type \textit{Decrease} since she looking for decreasing (lowering) causes of \textit{Glucose}.

To see if taking the regular insulin reduces Matt's glucose levels, Lisa selects the variable \textit{RegularIns} as a cause in Fig.~\ref{fig:glu-cdv}. The levels of \textit{RegularIns} are then rendered into the level bar chart. Lisa click the bars of \textit{high} and \textit{normal} to setup a value constraint on \textit{RegularIns}. Setting a 1 unit delay, the green (conditioned) distribution histogram is rendered on top of the unconditioned blue histogram. Since \textit{RegularIns} has many missing values (Matt only takes insulin three or four times a day), the green bars are much lower as a result. Lisa is delighted to see that the expected value of \textit{Glucose} is lowered from 143 to 122 and she clicks the 'Add' button to add [\textit{RegularIns}$\in$\textit{\{high,normal\}}] to the set of potential causes.

After following a similar procedure for [\textit{UltralenteIns=taken}] Lisa moves to the causal inference panel. The box configuration in the dashed inset of Fig.~\ref{fig:ins} shows the outcome of the analysis presented in Fig.~\ref{fig:glu-cdv} where the time delay was set to 1 hour and [\textit{RegularIns}$\in$\textit{\{high,normal\}}] and [\textit{UltralenteIns=taken}] are the two potential causes in \textit{X}. The inset shows that \textit{RegularIns} is a more significant cause than \textit{UltralenteIns} at that 1 hour time delay. However, the area chart in Fig. \ref{fig:ins} indicates that the max combined effect is reached after 4 hours. And so, by moving the time slider to 4, all visualizations in the panel update and we observe that \textit{UltralenteIns} now becomes the more significant cause. It essentially means that while both insulins are effective at lowering Matt's blood sugar level, \textit{UltralenteIns} brings a significant boost but this materializes only after some time.

Lisa now inspects the matrix chart to gain more insight on the relative strengths of these two insulin treatments. The diagonal tile of \textit{UltralenteIns} in the two matrices in Fig. \ref{fig:ins} always has a darker color than \textit{RegularIns} which verifies that it is a more impactful insulin treatment. However, with a 1 hour delay (matrix inside the dashed box), the tile at [\textit{c=RegularIns}, \textit{x=UltralenteIns}] is darker than that of [\textit{c=UltralenteIns}, \textit{x=RegularIns}]. This means the spurious influence of \textit{UltralenteIns} is removed in the significance test, making \textit{RegularIns} the more important cause for reducing \textit{Glucose}. This situation is reversed with the 4 hour delay, thus \textit{UltralenteIns} becomes the more significant one. It is also a darker tile indicating a larger influence.

All of this is quite enlightening to Lisa. To confirm what she has learned thus far she turns to the time sequence view to inspect the actual time series.  Fig. \ref{fig:ins-ts} shows a composite created from two screenshots when moving the mouse over the sequences where \textit{Glucose} and \textit{RegularIns} are in value mode and \textit{UltralenteIns} is in label mode. Under the 4 unit delay we set earlier and with the indicator lines at two time points, we observe that with the same dose of \textit{RegularIns}, the \textit{Glucose} would have a lower value (84) if \textit{UltralenteIns=taken} ('true', left time point), which confirms Lisa's prior findings. 

\vspace{+15pt}
\subsection{Temporal Causality Analysis: Air Quality dataset}

Suppose a public policy consultant, John, wants to study the cause of Shanghai's PM2.5 air pollution with DOMINO. As the first step he loads the Air Quality dataset and sets \textit{PMUSPost} to the \textit{Increase} type effect.

John soon realizes that exploring the potential causes one by one is tedious. So he queries DOMINO to obtain a first estimate. He knows that pollutants usually build up over time and to account for this he decides to try a time delay of 6 hours with the slider in the causal inference panel. Then he clicks \textit{Estimate Causes} and removes \textit{PMXuhui} from the causes since it is not a natural weather condition (\textit{P1}). The result is shown in Fig. \ref{fig:case-air}a 
-- high \textit{Temperature} ($>$15\textdegree C), low \textit{WindSpeed} ($<$25km/h) from the northwest or southwest (see \textit{WindDirection}, \textit{C} means windless), no \textit{Precipitation}, low \textit{Pressure} ($<$1016 pascal), and dry air (\textit{Humidity}$<$46\%) are found to be contributing to the build-up of PM2.5 pollution. If all these causes are satisfied, after 6 hours, the expected PM2.5 concentration would be almost twice the daily average (see the donut chart in Fig. \ref{fig:case-air}a). John further learns from the area chart that the build-up process lasts about 7 to 8 hours before reaching the peak.

One interesting observation here is that \textit{WindDirection} is more significant than \textit{WindSpeed}, which implies that external input by wind is a more important factor for Shanghai's air pollution (\textit{P2}). Fig. \ref{fig:case-air}b composes two screenshots of the time sequence view. The two highlighted indicators, as well as some other places along the sequence, show that while a strong wind coming from northeast (colored orange in the sequence of \textit{WindDirection}) is reducing PM2.5 concentration, the northwest wind (colored green) is not doing so. This makes perfect sense as next to Shanghai in the east is the Pacific Ocean while in the west is inland China. The wind coming from the sea brings clean, moist air (improving \textit{Humidity}), which reduces PM2.5 both directly and indirectly.

More details are revealed by the matrix view in Fig.~\ref{fig:case-air}a. The red tile in the \textit{WindSpeed} row and \textit{Temperature} column indicates that the latter event devalues the former's significance. The cell in the same row, in the column of \textit{WindDirection}, is also colored in very light color. Both indicate a temperature of [15, 20] degrees and certain wind direction are blocking the influence of low wind speed on air quality. 

Next, more insights are gained when John clicks the label \textit{WindDirection} in Fig. \ref{fig:case-air}b to visit the time delayed conditional distribution (\textit{P2,P3}). Fig. \ref{fig:case-air}c and d are the two distributions conditioning on only northwest or southwest wind. Comparing the two, it appears that the southwest wind, although occurring less frequently, had been bringing more pollutant, implying the major pollution source. At this point, John can already make some policy suggestions based on his findings. For example he might suggest starting a joint campaign with cities south- and northwest of Shanghai (such as Hangzhou and Nanjing) to curb air pollution. 

\textbf{Creating the causal flow chart.} 
John saves the current causal relations to the causal flow chart (see Fig. \ref{fig:overview}D) at time delay 5 (green links from \textit{Pressure, Temperature, WindDirection}, and \textit{Precipitation} to \textit{PMUSPost}). John then further explores the dataset by analyzing the chain effects among these factors. He first checks the cause of low \textit{Pressure} and discovers negative causal relations of \textit{WindSpeed} and  \textit{Temperature} (as well as \textit{WindDirection}) which are all well grounded in physics. These correct diagnostics give John confidence into the DOMINO software and he saves them to the causal flow chart in Fig. \ref{fig:overview}D (red links). Next, he checks the time delay between the pollution in \textit{PMUSPost} and \textit{PMXuhui} (southwest to the US embassy) caused by the wind. He learns that the time delay is relatively short which means any pollution measured in \textit{PMUSPost} is quickly felt in \textit{PMXuhui} and is solely determined by wind direction. It means that any alarm systems for excessive pollution must engage fast when the wind direction is unfavorable. He saves these relations to the causal flow chart as well.

The causal flow chart of Fig. \ref{fig:overview}D gives John an overview of the discovered relations. He can then revisit the saved results by loading the nodes (\textit{P4}) for further analysis, perhaps with new data collected after a policy change.

\begin{figure}[!t]
	\centering
	\subfigure[]{\includegraphics[width=3.45in]{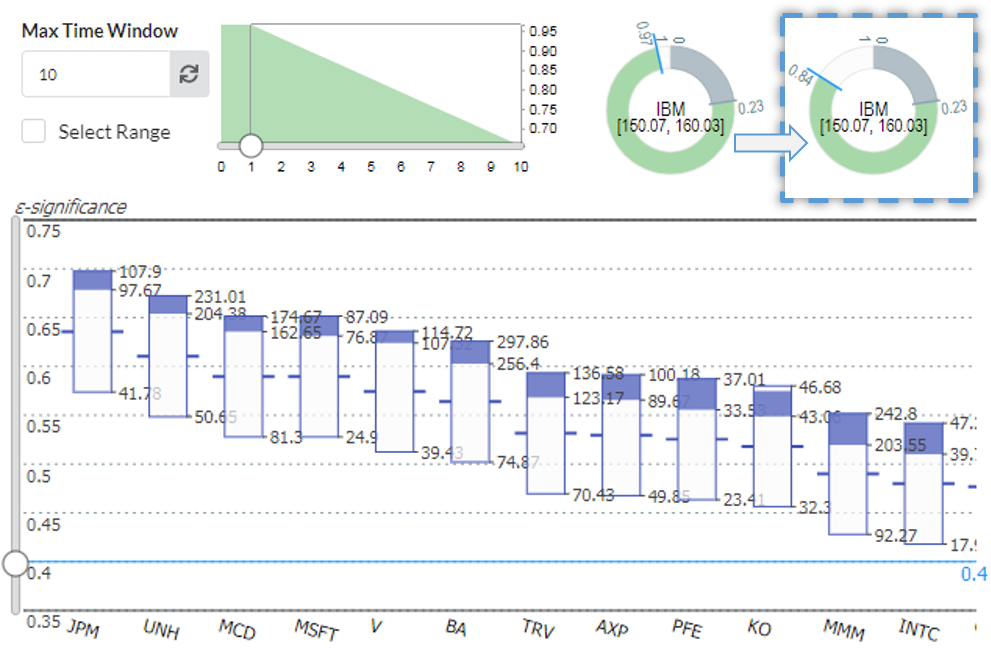}}
	\subfigure[]{\includegraphics[width=3.45in]{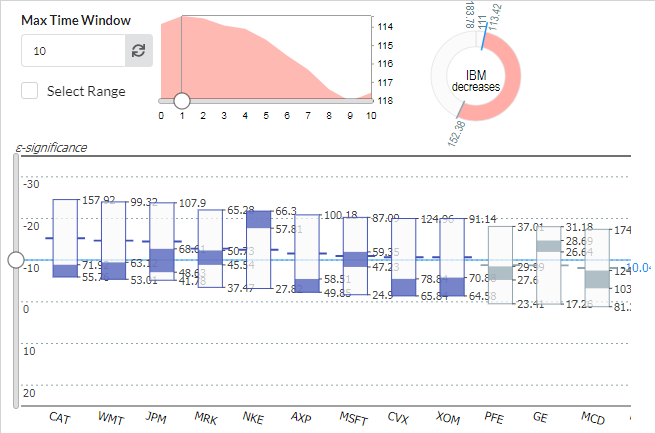}}
\vspace{-10pt}
  	\caption{Analyzing the DJIA 30 dataset. (a) Predictors of \textit{IBM} share price to fall into \$150-\$160 after 1 day. More boxes will render when dragging the chart. With the top five causes, the conditional probability drops from 97\% to 84\%. (b) Factors related to a decrease of \textit{IBM}'s share price.}
  	\label{fig:case-djia}
\vspace{-10pt}
\end{figure}

\subsection{Financial Strategizing: DJIA 30 Dataset}

In this scenario, a financial consultant, Jane, is serving a customer who wants to invest in IBM stocks. With five years of DJIA stock daily prices, Jane hopes to find out if there are any dependencies between the share price of IBM and that of others. These dependencies could be certain links between companies that have not yet been disclosed or are not obvious. Knowing such relations can
help the investor to (1) predict the development of prices of some specific stocks so that actions can be taken in advance, and (2) reduce the risk by investing in stocks that are not highly dependent.

Jane first wants to find out if there is any predictor for the share price of IBM falling into the range of 150 to 160 dollars, which is the target price range for the customer. A time window of exact 1 day is used as it is often believed that there is a sharp drop in influence after that time \cite{Kleinberg2010}. Loading the data, setting a \textit{ValueIn} type effect on \textit{IBM} with a value constraint of 150 to 160 in the conditional distribution view, clicking the auto-estimation button, and setting an  $\varepsilon$-significance of 0.4 yields the causal inference panel of Fig. \ref{fig:case-djia}. The visualizations are straightforward to understand -- when all causes occurred, i.e. the share prices of these stocks fell into the specified ranges, there was a 97\% chance of the desired effect to occur. However, to guide future actions using this result is somewhat impractical since the event of all predictor stocks falling into the desired ranges will be rare. To achieve a looser condition Jane slides the $\varepsilon$ threshold to a larger value. Then, with only the five most significant predictors in the display -- \textit{JPM} (JPMorgan Chase), \textit{UNH} (UnitedHealth), \textit{MSFT} (Microsoft), \textit{MCD} (McDonald's) and \textit{V} (Visa), the conditional probability drops to 84\% (the donut in the dashed area in Fig. \ref{fig:case-djia}a).

At the same time, there is also the possibility of the opposite effect, that is, stocks predicting the share price of \textit{IBM} to fall lower than its average. 
This case is investigated in Fig. \ref{fig:case-djia}b. 
Based on this visualization, we find that some stocks, e.g., \textit{CAT} (Caterpillar) fell under \$71.92 per share, \textit{WMT} (Wal-Mart) under \$63.12 per share, \textit{AXP} (American Express) under \$58.51 per share, etc., all predicting a low share price of \textit{IBM}. Thus it might be a good strategy to not buy them together with \textit{IBM} to lower the financial risk.

Due to space limitations, we cannot examine each of the stocks in the dataset, but doing so will likely lead to more valuable findings for financial strategy. Nevertheless, the presented case studies presented in this section show that DOMINO is well suited for the temporal causality analysis of data in drastically different domains.

\section{User Studies and Discussion}
To evaluate our work, we first interviewed four professors with online demos of DOMINO, and then performed interactive user studies with five junior scientists.
\subsection{User Feedback}
We engaged four professors (three male, one female) at our university, each native to a different scientific discipline: a climate scientist/ecologist (CS) with an emphasis on oceanography, a professor of public policy (PP) who studies the effects of technology on social inequality, a computational linguist/social scientist (CL) whose research seeks to understand health and well-being from large text data, and a psychology professor (PS) who is interested in the computational modeling of human behavior in decision making. All professors routinely perform data analytics to research causal relations in their domains. We opted for a demo-based evaluation since this promised the most feedback within the limited timeframe acceptable to these professionals. We demoed the pollution dataset since it had the most variety. Each of these one-hour sessions was conducted over a web conference with the first author performing the demo and the second author transcribing the feedback. The participants were encouraged to verbally steer the demo, ask for certain interactions, and pose what-if questions. The demo was conducted as a semi-structured interview where we sorted the feedback into the following four thematic categories (of questions over the course of each session):

\smallskip
\noindent
\textbf{Are the visual interface and the interaction defined on it intuitive? How do you rate their complexity?}
PS said that the system was complex but useful and intuitive. He said that it ``packs a lot in, but [that the] implicity is great" and he also noted that "there does not seem to be any waste of space." PP said that it ``makes sense, is easy to use, and visually intuitive." Everyone liked the amount of detail. However, all participants initially had trouble with understanding what information the donut chart encoded but a brief explanation cleared up these doubts in all cases. 

\smallskip
\noindent
\textbf{Do you think the system would be helpful in your work?}
All confirmed this at great fervor. CL said he ``could take the program right now and use it." Each participant then immediately went on to elaborate on some possible applications in their own work. For example, CL mentioned a use to track ``language across time to monitor the progression of depression." He said that ``medical psychiatrist could have good use for the software." TS said that he ``could see this to be helpful to look how countries vary over time [in their] technology use and adoption", asking, for example ``how cell phone use depends on GDP, or the other way around." 

\smallskip
\noindent
\textbf{What do you like most?}
Everyone liked the dashboard with the multiple views and the ``ability to interact [with it] in seemingly endless ways." PS commented that the ``Interface really allows people to dig in" and TS said that ``there are so many ways in which to look at the data." PS found that that ``being able to mess around with the latencies and time shifts" and ``tinker with the system" gave him a better understanding of the data. He especially appreciated the ability to ``automatically generate a solution and tweak it" and to gauge ``how well [his] own model and hypothesis meshed with the data." All participants especially liked the time series plot on the bottom right of the interface while the middle plot was reminding them of box and whisker plots which they were all very familiar with. It was also appreciated that our system could handle many variables. CS liked that our system supported categorical variables, which she said they ``often leave out because [they] do not use correlation." PS concluded that our interface succeeds in making ``causal inference much more accessible to people."

\smallskip
\noindent
\textbf{What do you think should be improved?}
One deficiency of our interface quickly exposed was that many of the axes were not labeled. CS mentioned that there should be support for spatial data. She also said that giving probabilities an event will not occur would be ``really valuable." CS and PS pointed out that at present our interface did not support interactions of multiple variables that could lead to a ``perfect storm." For example, it could be that ``temperature and wind direction each has no influence but together [might] have a lot of influence." CL and PS also asked for support for latent variables and unmeasured confounders. The treatment of outliers and drifts in data were also mentioned, both of which are presently relegated to a preprocessing stage. PS inspired us to ``expand the types of causal models" while TS asked whether our system could ``generate a report" stating that the ``nice graphics would be useful to export to a document."

\begin{figure*}[!t]
 \centering
 \includegraphics[width=7in]{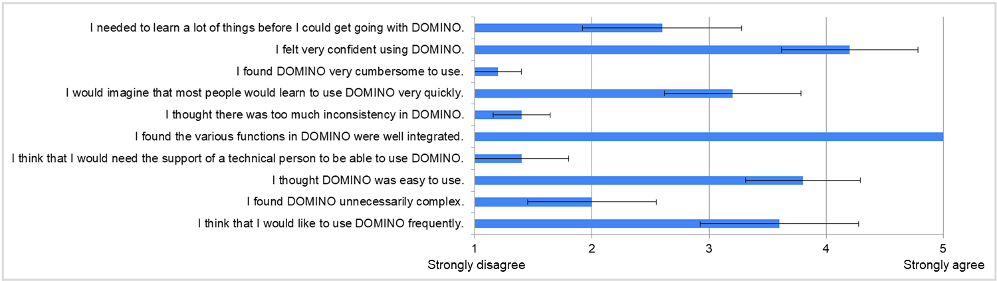}
 \vspace{-10pt}
 \caption{Results of SUS questionnaire with questions and summary statistics for the five participants. The bar calipers denote the standard error.}
 \label{fig:sus-qs}
 \vspace{-5pt}
\end{figure*}

\begin{figure}[!tb]
 \centering
 \includegraphics[width=3in]{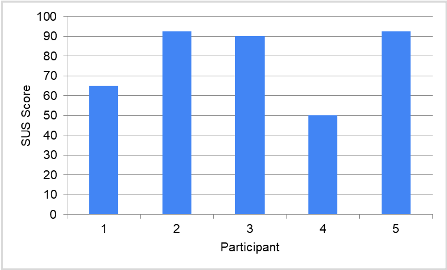}
 \caption{Aggregated SUS scores for all 5 participants. The scores range from excellent (3) to just acceptable (1) and nearly unacceptable (1). The average score is 78 (not shown), which is considered 'good'.}
 \label{fig:sus-sc}
\vspace{-10pt}
\end{figure}

Some rather fundamental comments came from PS who conducts research on how people use causality. He mentioned that ``people have rather complex notions about how things work, so if the machinery does not support those, people might be turned off." Users might have an internal model where they want to support ``negations and conjunctions, for example where wind speed is high but NOT from the northwest." These are complex causal relationships which our interface presently does not support. PS suggested to rather ``present [DOMINO] as a data exploration tool instead of a causal inference tool, a tool that can mine for causal relationships. This would lower expectations and be a better way to sell it." The outcome should be a collection of causal facts and not a promise that we ``find a causal model." He said that many people in science are ``regression happy" and that our program also does this ``but way better and in a more desirable way." He remarked that ``instead of multivariate regression, [our software] finds causal stories which would be very compelling to many people."

\vspace{-10pt}

\subsection{User Study}
We also conducted a formal user study with five junior scientists all in the final stages of attaining their PhD degrees. We chose junior scientists over senior ones, such as professors, since these are likely more engaged in hands-on active data analysis. We used a standard dataset to make the trials comparable and the effort practical. All but one participant were members of a special campus training program that aims to “provide STEM graduate students with interdisciplinary skills to translate complex data-enabled research into informed decisions and sound policies”. The campus program is designed to be diverse and so, in some sense, our study inherited that diversity. Specifically, the backgrounds of our participants were Technology, Policy and Innovation, Chemical Engineering, Operations Research, Climate and Atmospheric Science, and Computer Science. None of them had specific knowledge on causal inference but all were familiar with standard data visualization techniques.

Each session lasted about 1-1.5 hours and was conducted remotely over Zoom. We began by introducing our system using the Glucose dataset. We first demonstrated how one could arrive at certain insights via the manual mode; we then showed how this process can be accelerated using the automated mode and how one could further explore these outcomes. The participants could ask questions throughout the demo and all of them did so. Following, we put the participants into the driver’s seat. They were given a link to our server-installed software and shared their screen with us. We asked them to load the Air Quality dataset and gave them a quick intro into the dataset’s origins and variables. Then we asked them to examine the PM2.5 air pollution. 

We noticed that all participants needed some hand holding at first, but got used to the basic operations of the interface eventually. Most questions asked were about how the different views interacted with each other, and how to read the updated visualizations following each user operation. One of the most frequently asked questions was how the area chart updates if a different set of significant causes is selected after adjusting the threshold. Another confusion we often needed to clarify was how the two time sliders in the conditional distribution view and the causal panel worked differently. Eventually, all participants succeeded in arriving at conclusions similar to those demonstrated in Section 6.1.

Next, we allowed them to examine any other variable of their own choice. Most picked temperature or humidity and through interactions they learned that wind direction mattered as well as pressure. We observed that familiarity with the dataset's domain was most decisive on how quickly and accurately a participant could reach a conclusion.  Participants from Climate and Atmospheric Science could easily select possible causes for the increasing temperature even with manual selection, while others needed more time or had to rely more on automated estimation. We also found that all participants naturally followed the analytical process we established in Chapter 4. Very few questions about the interface were asked in this step, and so it seemed the participants had all gotten familiar with the system. At the exit interview all of them stated that they were confident about their findings. We also observed that the results they were able to obtain were quite similar to those we achieved in our own analyses, confirming good system utility.

After the session we gave each participant a link to a System Usability Scale (SUS) questionnaire. 
The SUS score is comprised of 10 questions each associated with a 5-level Likert Scale rating (1 for ‘Strongly Disagree’ and 5 for ‘Strongly Agree’). The questions and the statistics collected for each is presented in Fig. \ref{fig:sus-qs}. For the odd-numbered questions a 1-rating is best while for the even numbered questions a 5-rating is preferable. From the ratings we observe that the participants were overall satisfied with the system but felt that it took some effort to learn. We noticed that as well, but we also noticed that after our initial subtle assists they could nicely fly on their own and make independent discoveries.

Fig. \ref{fig:sus-sc} presents the aggregated SUS score for each participant. A score of 68 is at the 50th percentile range and considered acceptable. While there is no clear definition on score rating – the various publications are not conclusive -- it appears that scores greater than 85 are excellent while scores below 50 are not acceptable. Three participants gave an excellent SUS score, one gave a just acceptable, and one straddled the border of non-acceptable. This latter participant was a junior scientist who revealed during the session to be less data-driven and more qualitative. Nevertheless we found that the person was quite able to make some valid observations in the data but it did take significant effort. All others did much better, as their SUS scores also reveal. The average SUS score was 78 with a standard error of 8.7 which is considered ‘good’ and above the 80th percentile range.

\section{Discussion}
The user studies we conducted gave good insight on how analysts would use our system. They would begin with the manual exploration tools to gain a good sense for the nature and dynamics of the phenomena captured by the data. Upon gaining this familiarity, they would then switch to the automated estimation to add accuracy to these initial estimates. This would then form a starting point for a more thorough analysis using the various interactive tools.

\subsection{Limitations and Tradeoffs}

Causality is a complicated concept. Different theories may emphasize specific aspects and so do the systems built around them. This also applies to DOMINO which takes a probabilistic view where the inferred causation does not guarantee that causes always lead to the effect. While this is in line with some philosophical discussion of the nature of causality \cite{Eells1991,Williamson2009}, DOMINO might not fit in applications seeking deterministic relations. Questions like ``how much does MPG change if a car is 100kg heavier" cannot be answered with DOMINO, but might be modeled by other causality frameworks leveraging regressions \cite{Wang2016,Li2012}. But, on the other hand, DOMINO can better handle tasks where such probabilistic causality is preferred.

Secondly, DOMINO separates a cause from an effect entirely based on the time shift. Thus, to infer causality, DOMINO requires a multivariate time series which captures the delay a causal relation takes to act. If the data is non-sequential, or the time information is lost or averaged between rows, we can still process it by assuming a zero time delay. However, the results from applying the equations in Section \ref{sec:theory} will lose causal implication and become correlation. While the inferred relations can still be useful, their validity requires human knowledge to justify. 

Lastly, DOMINO focuses more on answering the question of ``why something happens" rather than achieving an overall understanding of the model governing the production of the observed data, i.e., a causal graph. While a user is able to gain a bigger picture of the model by continued application of the chaining relations and adding them to the causal flow chart,  constructing a complete causal model will require processing each time series as an effect, and then applying a causal inference method such as the PC algorithm \cite{Spirtes2000} to eliminate spurious edges.

\section{Conclusion}
We have established a workflow for exploring causality between time-dependent events. Implemented with logic-based theory, we presented a prototype system, DOMINO, that provides visual utilities for generating causal propositions and hypotheses and testing their truthfulness with time delays. Its effectiveness was demonstrated with two types of user studies. Future work will add more capabilities, such as support for interaction variables, categorical effect variables, and full causal graph evolution.


\section*{Acknowledgment}
This research was partially supported by NSF grants IIS 1527200 and 1941613.


%



\ifCLASSOPTIONcompsoc
\else
  \section*{Acknowledgment}
\fi


\ifCLASSOPTIONcaptionsoff
  \newpage
\fi



\bibliographystyle{abbrv}
\bibliography{references}
%



%
\vspace{-15pt}
\begin{IEEEbiography}[{\includegraphics[width=1in,height=1.25in,clip,keepaspectratio]{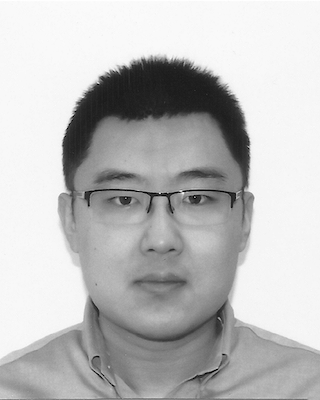}}]{Jun Wang} earned a PhD in computer science from Stony Brook University in 2018. This work is part of his dissertation. Jun also earned a BS and MS in computer science from Shandong University, China, in 2013 and 2009, respectively. He is now employed at Google in a R\&D position. Jun's research interests are visual analytics, information visualization, and data mining. For more information, see \url{https://junwang23.github.io/}.
\end{IEEEbiography}
\vspace{-15pt}
\begin{IEEEbiography}[{\includegraphics[width=1in,height=1.25in,clip,keepaspectratio]{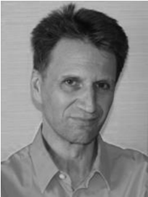}}]{Klaus Mueller} has a PhD in computer science and  is  a  professor  of  computer  science  at  Stony  Brook  University  and  a  senior scientist at Brookhaven National Lab. His research interests include explainable AI, visual analytics, data science, and medical imaging. He won the US NSF Early CAREER Award, the SUNY Chancellor’s Award, and the IEEE CS Meritorious Service Certificate. His 300+ papers were cited over 12,500 times. For more information, see \url{http://www.cs.stonybrook.edu/~mueller}.%
\end{IEEEbiography}







\end{document}